\input harvmac


%
%
\ifx\epsfbox\UnDeFiNeD\message{(NO epsf.tex, FIGURES WILL BE IGNORED)}
\def\figin#1{\vskip2in}
\else\message{(FIGURES WILL BE INCLUDED)}\def\figin#1{#1}\fi
\def\ifig#1#2#3{\xdef#1{fig.~\the\figno}
\goodbreak\midinsert\figin{\centerline{#3}}%
\smallskip\centerline{\vbox{\baselineskip12pt
\advance\hsize by -1truein\noindent\footnotefont{\bf Fig.~\the\figno:}
#2}}
\bigskip\endinsert\global\advance\figno by1}

\def\ifigure#1#2#3#4{
\midinsert
\vbox to #4truein{\ifx\figflag\figI
\vfil\centerline{\epsfysize=#4truein\epsfbox{#3}}\fi}
\narrower\narrower\noindent{\footnotefont
{\bf #1:}  #2\par}
\endinsert
}


\def\IC{{\ \hbox{{\rm I}\kern-.6em\hbox{\bf C}}}}
\def\IR{{\hbox{{\rm I}\kern-.2em\hbox{\rm R}}}}
\def\IZ{{\hbox{{\rm Z}\kern-.4em\hbox{\rm Z}}}}

\def\sIR{{\hbox{{\sevenrm I}\kern-.2em\hbox{\sevenrm R}}}}

\Title{RU-97-86, SU-ITP-97-26, UTTG - 25 -97, PUPT-1742}
{\vbox{\centerline{Schwarzchild Black Holes in Matrix Theory II}}}

\centerline{\it T. Banks~$^1$, W. Fischler~$^2$,  I.R. Klebanov~$^3$, 
L. Susskind~$^4$}
\medskip
\centerline{$^1$Department of Physics and Astronomy}
\centerline{Rutgers University, Piscataway, NJ 08855-0849}
\centerline{\tt banks@physics.rutgers.edu}
\medskip
\centerline{$^2$Theory Group, Department of Physics}
\centerline{University of Texas, Austin, TX, 78712}
\centerline{\tt fischler@physics.utexas.edu}
\medskip
\centerline{$^3$Joseph Henry Laboratories}
\centerline{Princeton University, Princeton, NJ 08544}
\centerline{\tt klebanov@puhep1.princeton.edu}
\medskip
\centerline{$^4$Department of Physics}
\centerline{Stanford University, Stanford, CA 94305-4060}
\centerline{\tt susskind@dormouse.stanford.edu}

\bigskip

\medskip

\noindent

We present a crude Matrix Theory 
model for Schwarzchild black holes in uncompactified
dimension greater than $5$.  The model accounts for the size, entropy,
and long range static interactions of black holes.  The key feature of
the model is a Boltzmann gas of D0 branes, a concept which depends on
certain qualitative features of Matrix Theory which have not previously
been utilized in studies of black holes.
\hyphenation{Min-kow-ski}
\Date{October 1997}

\newsec{\bf Introduction}

In this paper we will present a crude and simple quantum mechanical
model of Schwarzchild 
black holes in noncompact spacetimes of dimension $11 \geq D
\geq 6$.  The model accounts for many of the gross properties of a black
hole: its mass, entropy, radius, as well as the Newtonian gravitational
interaction between two black holes .  We hope that
future investigations will provide more evidence that it indeed describes
black holes. 

The model is based on the recent Matrix Theory 
\ref\bfss{T.Banks, W.Fischler, S.Shenker, L.Susskind, Phys.Rev. D55
(1997) 112, \break hep-th/9610043. } proposal for
a nonperturbative, light cone frame formulation of M theory.
This proposal is far from complete.  In particular, it has proven
extraordinarily difficult to find a general description of Kaluza-Klein 
compactification of Matrix Theory.  We will try to formulate our
model in a way which uses as few of the detailed properties 
of Matrix Theory compactification as possible.

The essence of our black hole model is simplicity itself: {\it for the
most part, the black hole consists of a Boltzmann gas of D0
branes\ref\jopo{J.Polchinski, 
{\it TASI Lectures on D-Branes}, hep-th/9611050  } \foot{After
submitting this paper, we were informed by I. Volovich of a previous
paper in which black holes are treated as a Boltzmann gas of D0 branes, 
\ref\volo{I.V.Volovich, Talk given at 2nd International Sakharov
Conference on Physics, Moscow, Russia, 20-23 May 1996, hep-th/9608137.}}

 interacting 
via the long range (IMF)-static forces which are
associated with graviton exchange with zero longitudinal momentum.}
The phrase \lq\lq for the most part\rq\rq\ means that the bulk (at least
a finite fraction) of the entropy and energy of the black hole is
accounted for by the D0 brane gas.  Furthermore, the Schwarzchild radius
of the hole is determined by the quantum wave functions of the D0
branes.  The other constituents of the hole lie within its Schwarzchild
radius. We will also show that the long range static Newtonian
interaction of two equal mass black holes (the technical reason for
considering only equal masses will be explained below) can be completely
understood in terms of the velocity dependent forces between D0 branes
and their velocity distribution inside the black hole.

We have mentioned other constituents of the black hole.  The necessity
for such follows from our insistence on a {\it Boltzmann} gas of zero
branes. D0 branes are bosons (and fermions).  Their statistics symmetry
is the residuum of a continuous gauge symmetry which is restored when
they are very close together.   We will see that a Bose or Fermi gas
does not have enough entropy to account for the properties of black
holes.  The Boltzmann nature of the D0 branes inside a black hole
derives from the existence of a {\it classical background} configuration
of the matrix model.  The requisite D0 branes are a certain class of
matrix fluctuations around this background.  The background completely
breaks the gauge invariance of the model, including the would be
statistics symmetry of the zero branes.  In a picturesque language which
we will explain below, the zero branes are {\it tethered} to the
background in a way which makes them distinguishable.  

The background must satisfy a number of properties in order for the
model to make sense.  It is in trying to verify the existence of
appropriate backgrounds that we must appeal to our rather imperfect
knowledge of the detailed dynamics of the compactified matrix theory. 
We have been able to make progress on this problem only when $D = 11$,
or $8$ and we have some partial results for $D=7$.  Furthermore,
beginning with $D=5$ there seems to be a serious problem with our ideas,
and when $D=4$ the logarithmic nature of the transverse interactions
between D0 branes presents us with some unresolved questions.  In order
to clearly separate these issues from the central simplicity of our
model, we have decided to present things in the following order.  In the
next section, after recalling some facts derived in a previous paper,
\ref\bfks{T.Banks, W.Fischler, I.R.Klebanov, L.Susskind, hep-th/9709091.} , 
we use the Boltzmann gas model to derive some of the
properties of black holes.  We then present the general criteria for
backgrounds which could justify the Boltzmann gas model, and present
examples of such backgrounds in $11$ and $8$ dimensions.  Finally, we
explain the difficulties in $D= 4,5$.   Our approach throughout is based
on a rule of thumb motivated in \bfks\ which suggests that for the study
of Schwarzchild black holes, the optimal value of $N$  (the value which
is large enough to obtain an adequate description without involving many
redundant variables) is of order the entropy, $S$, of the black hole. 
 In the appendix we discuss some
highly speculative arguments which suggest that the appropriate Matrix 
Theory setting for the study of black holes in $D=4,5$ and perhaps even
$6$ is the regime where $N \gg S$.

\newsec{\bf The Boltzmann Gas Model}

We begin by recalling some results of \bfks\ .  Our basic strategy is to
work in the light cone frame with a compactified lightlike direction of
circumference $R$ and
total longitudinal momentum $N/R$.  As we increase $N$, the number of
degrees of freedom in the model increases.  The optimal choice of $N$
for describing any given system depends on the system.  If we choose $N$
too small we do not describe the system adequately, while if we choose
it too large we include many redundant variables which are frozen into
their ground state in the system under consideration.  Our criterion for
fixing the optimal value of $N$ for a system described by a classical
spacetime geometry is that the geometrical size of the system, $R_S$, must
, after a longitudinal boost to momentum $N/R$, fit into the
compactification radius. 
This gives the equation
\eqn\nfit{{M R \over N} R_S = R}
or
\eqn\nfitb{N = M R_S}
where $M$ is the mass of the system.  This is the minimal value of $N$
which can give us an adequate description of the system, and larger
values of $N$ give us a description with many redundant variables.
 For a generic system there is an as yet
unresolved ambiguity about which geometrical parameter $R_S$ to use in
this equation, but for Schwarzchild black holes $R_S$ is clearly the
Schwarzchild radius, since there is no other length scale in the
Schwarzchild geometry.  In this case it is easy to see that $N$ is
the entropy of the black hole (all such statements are meant as order of
magnitude estimates, no precise numerical coefficients are computed in
this paper).

The other result of \bfks\ which we wish to recall is the description of
Hawking particles in the light cone frame.  In the rest frame, a typical
Hawking particle has energy $T_H \sim M^{-{1\over D - 3}}$ 
and isotropically distributed momenta
of the same order of magnitude.  After the boost it has longitudinal
momentum ${ N \over MR} T_H \sim 1$.
Its transverse momentum is of order $T_H$ and its light cone energy is
of order ${MR \over N} T_H \sim N^{- {2\over D-2}}$.  
We will see that these are precisely the
kinematic properties of one of our Boltzmann D0 branes.

We are now ready to present our model.
For concreteness, we restrict attention to toroidally compactified
matrix models.  We believe that our considerations will turn out to be
more general than that, since they rely only on very general properties
of the model.   In particular, in all known versions of Matrix Theory,
momentum in the noncompact directions is carried only by $DN$ degrees of
freedom  $X^i_a$ which represent the positions of $N$ D0 branes in the 
noncompact dimensions.  Due to the gauge invariance of the underlying
model, there is some ambiguity about how these D0 brane positions are
extracted from the full set of degrees of freedom.  The role of the
background, which we discuss in the next section, is to resolve this
ambiguity. For the moment, we simply assume that the $X^i_a$ can be
thought of as the coordinates of distinguishable particles.  

When the D0 branes are far apart, one can calculate an effective
interaction between them \ref\dkps{M.R.Douglas,
D.Kabat, P.Pouliot, S.H.Shenker, Nucl.Phys. B485 (1997) 85,
hep-th/9608024.} which has the form (if the D0 brane
velocities are slow, which we will verify self consistently in a
moment):
\eqn\effham{ H = \sum_{a=1}^N {\bf p}_a^2 + A G_N \sum_{a,b} {({\bf p}_a -
{\bf p}_b)^4 \over \vert {\bf X}_a - {\bf X}_b \vert^{D-4}}}
When $D=4$ the power law interaction is replaced by a logarithm. $G_N$
is the dimensionally reduced Newton constant, and the coefficient $A$
was calculated in \dkps\ .
This is indeed the correct interaction between uncompactified gravitons
in $D$ spacetime dimensions.  In Matrix Theory it is calculated by
integrating out the Super Yang Mills degrees of freedom other than the
D0 brane coordinates.   

Now consider a metastable configuration in which all $N$  D0 branes are bound 
together.  In the spirit of the mean field approximation, each $D0$
brane should have approximately the same mean distance $R_S$ from the
center of mass, and (by the uncertainty principle) the same mean
momentum $1/R_S$.  Equating the kinetic and potential terms in the
Hamiltonian of the bound system (the virial theorem), we obtain
\eqn\radius{B G_N^{-1} R_S^{D-2} = N}
The numerical constant $B$ cannot be calculated with our present crude
methods.  We are treating the system of D0 branes as a Boltzmann gas.
Thus its entropy is of order $N$.  We have thus derived the
Bekenstein-Hawking area law (up to a numerical constant) from the basic
principles of Matrix Theory.\foot{There is a subtle loophole in this
argument which we will return to when we discuss backgrounds in the next
section .}\foot{As this paper was in preparation, we received  preprints
by Horowitz and Martinec \ref\ejm{G.T.Horowitz, E.J.Martinec,
hep-th/9710217.} and by Li \ref\li{M.Li, hep-th/9710226.}, which also
derive the area law in
arbitrary dimensions.  These authors also claim to calculate the entropy
as order $N$, but do not make explicit the fact that one must treat the
D0 branes as distinguishable.}

We can also calculate the energy of the system to the same accuracy.
The single particle energies are of order
$R_S^{-2}$, so the total energy scales like $N^{1 - 2/(D-2)}$.  Using
the standard relation between light cone energy and mass, we find that
the mass of the system scales like
\eqn\mass{M \sim G_N^{-1/(D-2)}
N^{1 - 1/(D-2)} = G_N^{-1/(D-2)} N^{(D-3)/(D-2)}}
Again, the dependence on Newton's constant follows from dimensional
analysis and the fact that the radius, $R$, of the lightlike circle is
only an overall scale in the Hamiltonian.  Thus, the correct
relation between mass and Schwarzchild radius also follows from our
model. 

Note that the kinematics of the individual D0 branes (transverse
momentum of order $R_S^{-1}$, longitudinal momentum of order one), is
precisely that of boosted Hawking particles.  Thus it is tempting to
identify the elementary process of Hawking radiation as the emission of
a single D0 brane from the bound cluster.  In order to discuss the rate
of this process, we will have to say something about the nature of the
backgrounds.    Another consequence of these kinematical relations is
the necessity for treating the D0 branes as a Boltzmann gas.  The energy
per particle, or effective temperature, is of order $R_S^{-1}$ and goes
to zero with $N$.  In a low temperature Bose or Fermi gas, the entropy
per particle vanishes as a power of the temperature.   Thus, we must
assume Boltzmann statistics for the zero branes in order to reproduce
the entropy of black holes.

As a final application of our model, we will compute the Newtonian
gravitational potential between a pair of equal mass static black
holes\foot{The idea for this calculation is due to S.H.Shenker, who also
collaborated in the actual computations.}.
The restriction to equal mass is a consequence of our strategy of 
choosing an optimal value of $N$ to study any system.  Black holes of
different masses have very different numbers of excited degrees of
freedom.  We cannot give an adequate description of the combined system
without \lq\lq overdescribing\rq\rq\ the black hole of smaller mass.
We have yet not understood how to study such a system in the matrix model.

Another important qualification is that present Matrix Theory technology
restricts us to the study of processes with zero longitudinal momentum
transfer. Thus rather than studying black holes at fixed positions in
their common rest frame, we must average the potential over the
longitudinal circle.  This introduces an interesting subtlety into the
calculation: the length of the averaging circle in the boosted frame is
Lorentz contracted relative to that in the rest frame.  We will do a
calculation of the light cone energy of the two black hole system in the
boosted frame.  We then use the usual connection between light cone
energy and rest mass to calculate the rest energy, and thus the static
potential of the system.  This must be interpreted as the Newtonian
potential averaged over a longitudinal circle whose size is related to
that in the boosted frame by a Lorentz contraction factor.

To compute the light cone energy of a pair of black holes, we take into
account only the velocity dependent interactions between their zero
brane constituents:
\eqn\inta{\delta P^- =  A G_N \sum_{a,b} {({\bf p_1}_a -
{\bf p_2}_b)^4 \over \vert {\bf X_1}_a - {\bf X_2}_b \vert^{D-4}}}
This should be averaged over the velocity distribution of the two holes.
Many terms drop because of spherical symmetry of the distributions and
we are left with
\eqn\intb{\delta P^- \sim N^2 {G_N \over r^{D-4}} ({G_N N})^{- 4 \over
(D-2)}}
where we have used $r$ to denote the distance between the centers of
mass of the two black holes and inserted our result for the average
velocity. 
This is to be compared with
\eqn\intc{\delta P^- = {{(2M + V(r))^2 \over N} - 4 {M^2 \over N}} =
8{M\over N} V(r)}
We obtain
\eqn\intd{V(r) \sim G_N {N^3 \over M r^{(D-4)}} ({G_N N})^{- 4 \over (D-2)}}
Using the Bekenstein-Hawking mass-entropy relation (which we have
derived above), we get:
\eqn\inte{V(r) \sim G_N {M^2 \over r^{(D-4)}} 
\left (G_N N \right)^{-{1\over (D-2)}}} 
This is the correct formula for the static Newtonian potential
appropriately averaged over the longitudinal circles.
To see this note that the last factor is $1/R_S$. Indeed, 
the averaging should introduce a factor of $1/R$, and if we are
in the rest frame of the black holes and work with $S\sim N$,
then $R\sim R_S$.

An important question is whether the more complicated multibody forces
which occur at higher orders in the matrix model perturbation expansion
can effect the calculations which we have done.  For example, the three
body interaction computed by Dine and Rajaraman\ref\dr{M.Dine,
A.Rajaraman, hep-th/9710174. } scales like
$N^3 {v^6 \over r^{14}} $ in eleven dimensions.  Note that for values of
$v \sim {1\over r}\sim N^{-(1/9)}$ this has the same order of magnitude
as the terms we have included in our analysis\foot{This observation is
due to S.H.Shenker, who informs us that the general term in the
perturbation expansion obeys this scaling rule. It is of the form
$N^k {v^{2k} \over r^{7(k-1)}}$.}.  Thus, although a
quantitative analysis of the black hole will have to take into account
all sorts of complicated multibody interactions, the scaling laws which
we have derived will still be obeyed.

\newsec{\bf Background Information}

We now turn to the central question of why it is proper to treat D0
branes as Boltzmann particles.  The basic intuition is that, in the
matrix model, particle statistics is part of a larger gauge group, which
acts on degrees of freedom in the theory which are not particle like.
They do not carry momentum in the spacetime of the matrix model, but may
be viewed as internal excitations of the partons which do carry such
momentum. In some sense, this is one of the key feature which distinguishes
Matrix Theory from quantum field theory (the other is its holographic
nature).  

Consider a background 
configuration , $M_{cl}$ of Matrix Theory described by some fixed value
of the matrix variables, $M_{cl}$.  In particular, it will have fixed
values ${\bf X}_{cl}$ of the position variables whose zero mode is
shifted by a multiple of the unit matrix under spatial translations.
The background must satisfy a number of properties:

\item{1.} It must be a metastable, classical configuration of the matrix
model, {\it i.e.} the center of a coherent state with long lifetime.

\item{2.} It must not be left invariant by any element of the gauge group. 
As a consequence of the first postulate, this leads to a situation in
which the gauge group is \lq\lq spontaneously broken\rq\rq .  Since we
are always dealing with a finite quantum system, this means that the
time scales for motions of the collective coordinates which gauge rotate
the classical background are much longer than any of the other time
scales in the system.

We then define a configuration of \lq\lq zero branes in the classical
background $M_{cl}$\rq\rq\ as the configuration in which we shift the position
coordinates by
\eqn\zeroshift{{\bf X}_{cl} \rightarrow {\bf X}_{cl} + \sum_{a=1}^N {\bf
x^a} \delta_a ,}
where $\delta_a$ is a collection of $N$ commuting $N\times N$ matrices
which we will specify more precisely in a moment.

All compactifications of Matrix Theory whose Hamiltonian is understood
contain a term
\eqn\comm{- Tr\ tr\ [X^i, X^j]^2}
where the $X^i$ are related to coordinates in the noncompact space
dimensions. They are $N \times N$ matrices whose matrix elements are
operators in another Hilbert space.  The small $tr$ refers to the trace
in this internal space, while $Tr$ is the matrix trace.   
Such a term will give rise to harmonic potentials for the coordinates
${\bf x^a}$ in the classical background ${\bf X_{cl}}$.   For a given
classical background, we will choose the $\delta^a$ in such a way as to
minimize the coefficients of these harmonic potentials.  We will give
examples of this below in particular dimensions, and argue that it is
always possible to make these terms as small as ${1\over N} \sum {\bf
x^a}^2$.  If the dimension $D > 5$ then it is easy to see that, for
distances of order $R_S \sim (G_N N)^{1/(D-2)}$, the
harmonic energy is no larger than the energies associated with zero
brane interactions\foot{Note that in principle we should rederive the
two body zero brane interactions in the presence of the background, by 
calculating the one loop quantum fluctuations around the configuration
\zeroshift\ .  Since the background is supposed to receive only small
quantum corrections itself, it seems obvious that for large ${\bf x^a}$
the result will be the same as that in the absence of the background,
but this point should be checked.}
 (in $D=6$ the two energy scales are the same).  

We will impose two more constraints on the classical background:

\item{3.} The total energy of the background configuration should be
smaller than or equal to the total zero brane energy $N^{1- {2\over
D-2}}$ in dimensionally reduced Planck units.

\item{4.} The classical size of the background, should be smaller than
or equal to the Schwarzchild radius.  We will comment on this condition
further below.

If we can find backgrounds which satisfy all of these conditions then we
can remedy the difficulties of the zero brane model of black holes which
we presented in the previous section.  That model actually had two
inconsistent features.  The first of these was the treatment of the zero
branes as if they were distinguishable particles.  This is resolved by
the existence of the background.  Mathematically, the configurations
described by Equation \zeroshift\ have no residual $S_N$ gauge symmetry
which could act as particle statistics.  Intuitively, the interactions
with the background distinguish the zero branes from each other.  We
will see a particularly vivid physical picture of this in the eleven and
eight dimensional examples.

The second unclear feature of the model of the previous section 
was the assumption of a
bound configuration of zero branes.  The matrix Hamiltonian has only one
exact bound state configuration - the supergraviton with longitudinal
momentum $N$.  This is a threshold bound state.  Every other state in
the theory eventually breaks up into separated asymptotic supergraviton
states (the asymptotic supergravitons with longitudinal momenta smaller
than $N$ are also
stable excitations of the model, but they do not correspond to a bound
configuration of $N$ zero branes).  
The background makes explicit the necessity of finding metastable
excitations of the system.  Once the metastable background configuration
is present, its zero brane excitations are bound to it by harmonic
forces.  It is only via the quantum processes which allow the background
to fluctuate that the system can decay.

The model of metastable classical background plus zero branes predicts
the correct relation between the mass and Schwarzchild radius of a black
hole, as well as the gravitational interaction between black holes of
equal mass.  We would like to claim that it also predicts the correct
Bekenstein-Hawking mass entropy relation for black holes.  In order to
do this we must understand the entropy coming from summing over
background configurations.  It should turn out to be less than or equal
to the zero brane entropy.  We do not have a general or rigorous understanding
of why this is so.  A complete answer to this question would constitute
a proof of the Bekenstein bound on entropy in the context of the matrix
model (in \bfss\ it was argued only that the bound was satisfied by the
ground state wave function).  We do not have such a proof.  In
dimensions $11$ and $8$ we believe that we understand the entropy
of background configurations satisfying the conditions outlined above.
This shows that the configurations which we study do satisfy the entropy
relation for black holes.
We do not have an argument that configurations other than the ones we
have studied cannot pack larger entropy into the same area.

Another issue is raised by a discussion of the entropy of backgrounds. 
Any configuration satisfying the criteria we have outlined will behave
like a black hole.  But what is a typical background?  In particular, 
if there are
background configurations with entropy of order that of the zero branes,
then they must be understood if we ever wish to compute the coefficient
in the Bekenstein-Hawking formula.
We cannot make any more progress on these questions without turning to
particular examples.  

\newsec{\bf Schwarzchild Black Holes in Eleven Dimensions}

In eleven noncompact dimensions, it is natural to search for the
semiclassical backgrounds of our black hole model among the membrane
excitations of the eleven dimensional matrix model.  The latter are
semiclassical excitations of the model whose quantum metastability is
guaranteed by their closeness to exact BPS configurations \bfss\ 
\ref\bss{T.Banks, N.Seiberg, S.Shenker,Nucl.Phys. B490 (1997) 91,
hep-th/9612157. }.  
Let us remind the reader of how these configurations are constructed.
The matrices
\eqn\compset{U^k V^l} where 
$U$ and $V$ are unitary matrices satisfying $U^N =1 = V^N$, and 
\eqn\comm{UV = e^{{2\pi i \over N}} VU ,}
are a complete set of $N\times N$ matrices.  For large $N$ we
concentrate on
those matrices whose expansion
\eqn\exp{M = \sum M_{kl} U^k V^l}
approaches the Fourier expansion of a smooth function on a torus with
coordinates $p$ and $q$ in the interval $[0,2\pi ]$.  The identification
is made via the formal substitution $U \rightarrow e^{iq}, V\rightarrow
e^{ip}$.   For such matrices commutators approach ${2\pi i \over N}$
times the Poisson bracket of the corresponding functions, and the trace
approaches $N$ times the integral over the torus.  
The matrix model action transmutes under these formal substitutions into
the action of the light cone supermembrane \ref\dhn{B. de Wit, J.
Hoppe, H. Nicolai, Nucl.Phys. B 305 [FS 23] (1988)
545. }.  

A membrane configuration is simply a choice of the ${\bf X}$ coordinates
of the matrix model to be such smooth functions on the torus\foot{We
choose the torus only for simplicity of exposition.  Actually for
sufficiently large $N$ we can
find matrices reproducing membranes of any finite genus with any stated
accuracy , and in principle any
such background will serve the same purpose.  For fixed $N$, only a
finite number of genera will be well approximated.}
A generic such choice will break all of the $U(N)$ gauge symmetry.  
Furthermore, it is clear from the construction that for a large smooth
membrane these are long lived semiclassical states.

Now consider the matrices defined by the following periodic Gaussian
functions 
\eqn\obranedef{\Theta_{p_i, q_i} = \sum_{k,l} e^{ - N[(p - p_i - 2\pi
k)^2 + (q - q_i - 2\pi l)^2]} .}   
For large $N$, the commutator of two of these matrices is well
approximated by the Poisson bracket and is smaller than
\eqn\bound{e^{- \ha N [(p_i - p_j)^2 + (q_i - q_j)^2]}}
Taking a distribution of points $(p_i, q_i)$ spaced by distances
of order $1/\sqrt{N}$ we can obviously make this commutator as small as
we like.  There are $o(N)$ such points and so we can find $o(N)$ matrices
$\delta_a$ of this form.

The $\delta_a$ are normalized to $Tr \delta_a = 1$
They are the \lq\lq cells in phase space\rq\rq\ which make up individual
D0 branes.  Note that we have implicitly used the classical membrane
configuration to define the $\delta_a$ .  We cannot do unitary
transformations on $\delta_a$ which leave the membrane configuration
unchanged. Thus, there is no statistics symmetry as there is for D0
branes in free space.
Each zero brane is \lq\lq tethered\rq\rq\ to some particular
point in the membrane volume.  

We can now compute the harmonic potential between the zero branes and
the membrane by plugging the configuration $X_{cl} + \sum x^a \delta_a $
into the membrane Hamiltonian.  The terms bilinear in {\it different}
$\delta a$ are of the same order as the commutator between different D0
branes, and so we drop them.  The total harmonic potential is thus
\eqn\harmpot{\delta H \sim \sum ({\bf x^a})^2 N {\int} dp dq\ 
[p^2 + q^2]({\nabla X_{cl}})^2 e^{- N(p^2 + q^2 )}.}
The factor $N$ in front of the integral comes from the $1/N$ in the
definition of the energy in terms of membrane variables, and two factors
of $N$ originating in the derivatives of the $\delta_a$.  For truly
smooth membranes, the gradient of $X_{cl}$ is independent of $N$ as
$N\rightarrow \infty$.  The harmonic potential then has an overall
coefficient $1/N$.  It will also be useful to consider configurations
$X_{cl}$ whose gradients scale like some positive power of $N$.  
In principle, if we wanted to use these in the construction of black
holes, we would have to investigate their stability properties, since it
is no longer clear that they are close to being BPS states.
In the end, we will discard such configurations for other reasons so we
will not stop to perform this investigation.
The harmonic potential for such a configuration is larger by a factor of
$k^2$, where $k$ is its maximum wavenumber, than that for a smooth
membrane. In any event, we must have $k < \sqrt{N}$.  For larger values
of $k$ the membrane description of matrices is completely wrong.  There
are no matrices corresponding to such short wavelength membrane
configurations. 

We have now essentially completed the construction of a black hole model
in eleven dimensions.  The energy of smooth membrane configurations is
of order $1/N$ and thus negligible compared to the energy $N^{7/9}$ (or
even to the energy per particle, $N^{-2/9}$).  The harmonic potential
experienced by a single zero brane at the
Schwarzchild radius $R_S \sim N^{1/9}$ is of order $N^{- (7/9)}$, also
much less than the energy per particle of the gas of interacting zero
branes. Finally, the entropy of all possible smooth membrane
configurations can be estimated as that of a cutoff $2+1$ dimensional
quantum field theory, with cutoff independent of $N$.  The entropy is
thus of order $1$\foot{For each membrane we can sum over all
configurations of the D0 branes, so the membrane and zero brane
entropies should be added.} and makes a negligible perturbation to the
zero brane entropy $N$.   

We can try to find a larger set of configurations by allowing $k$ to
grow with $N$.  The condition that the harmonic potential not interfere
with the interacting zero brane dynamics is $k^2 < N^{5/9}$.  The
membrane entropy is of order $k^2$ (volume of the field theory in cutoff
units), and is still negligible compared to the zero brane entropy.
In order to have entropy of order $N$ from membrane configurations, we
would have to consider membranes with a wave number cutoff of order
$\sqrt{N}$ .  This is the edge of validity of the membrane picture \foot{It
should be clear that such configurations could not be taken as backgrounds
in our model.  We are discussing them merely as an example of high
entropy configurations which are not contained within the 
Schwarzchild radius.}.  These configurations have energy at least as
large as $N$ and do not satisfy the black hole mass-entropy relation.
Their mass is at least $N$, larger by a factor $N^{1/9}$ than that of a
black hole of similar entropy.  The
Schwarzchild radius of a black hole of this size will not fit into the
compactification volume.  It is not easy to obtain a first principles
 matrix model estimate
of the size of these configurations.  A normal $2+1$ dimensional field
theory estimate of the mean square fluctuation of the membrane
coordinate gives $<x^2>^{\ha} \sim N^{1/4}$.  This is much larger than
the Schwarzchild radius of a black hole of mass $N$ in $11$ dimensions.
We suspect that, if anything, this is an underestimate of the fluctuation
of the membrane coordinate in this cutoff, nonrenormalizable field theory.

Thus, among membrane configurations of the matrix model, none can be
found which compete with the membrane plus zero brane gas in putting a
large amount of entropy into a small area.  If we could demonstrate the
same thing for all other states of the matrix model, we would have
proven the Bekenstein bound in eleven dimensions.

\newsec{\bf Eight Dimensional Black Holes in Matrix Theory}

We have studied the system with eight noncompact dimensions in \bfks\ . 
Here we elaborate the discussion and take a somewhat different point of
view.  Our general approach in the current paper suggests that we look
for a classical background to distinguish the zero branes.  We can of
course study the same type of background which we used in eleven
dimensions - membranes.  However, we will find a much more common type
of background in the eight dimensional version of the theory, one which
contributes a finite fraction of the energy and entropy of the black
hole.  Indeed, the typical background is described by the configurations
we studied in \bfks\ .

In order to have entropy of order $N$ and carry a finite fraction of the
energy, the background must consist of $o(N)$ independent excited
degrees of freedom, each carrying an energy of order the temperature,
$N^{- (1/3)}$.  This means that most of the degrees of freedom of the
$3+1$ dimensional SYM theory are irrelevant for these considerations,
for they carry energy of order one.  The relevant DOF are those of the
uncompactified model, except that the coordinates in the compactified
directions are angle variables, since they originate as Wilson loops
of the $3+1$ gauge fields.  The first component of the background is a
classical expectation value of these Wilson loops corresponding to a
more or less uniform lattice of D0 brane positions on the three torus,
with spacing $N^{-(1/3)}$.  This background breaks the gauge symmetry
down to $U(1)^N$, the subgroup preserving the basis in which the Wilson
loop is diagonal.  Note in particular that there is no permutation gauge
symmetry among the eigenvalues.  Each eigenvalue is labelled uniquely by
its position on the three torus.  The energy of this configuration of
Wilson loops is of course zero.  We will call off diagonal matrix
elements in the basis in which the Wilson loop is diagonal, {\it charged
fields} .  Those corresponding to gauge field components will be called
W bosons, and those corresponding to noncompact position coordinates
will be called charged Higgs fields.

In the presence of the Wilson loop, the charged W boson and Higgs fields
(the components of off diagonal matrices in and perpendicular to the
three torus respectively) 
have no zero energy modes.  The $IJ$ matrix element\foot{We use
multiindex notation.  Capital letters refer to triplets of integers
specifying positions on the zerobrane lattice.}  feels a potential
$A_I - A_J$ .  For points separated by $o(1)$ links of the lattice, this
is of order $N^{-(1/3)}$.  Of course this is also the energy of the
lowest lying mode of this variable.  Thus, ignoring for the moment the
term in the Hamiltonian quadratic in commutators of the charged fields
with each other, we should be able to excite each of 
these degrees of freedom to
a classical state with energy of order the temperature.  Thus, for each
D0 brane we excite $o(1)$ of these degrees of freedom (its links to a
few nearest neighbors).  This gives the background an entropy of order
$N$ - there are $e^{cN}$ different backgrounds with the same characteristics.

The zerobrane positions in the noncompact directions are the diagonal
matrix elements ${ r^a}$ (with $1\leq a\leq 6$) 
of the Higgs fields.  Note that because of the nature of
the Wilson loop there is no residual $S_N$ gauge symmetry permuting
these variables.  The harmonic potential for these variables 
has the form
\eqn\harmpot{<\vert W_{IJ}\vert^2 >( r_I^a - r_J^a )^2 .}
where the brackets denote averageing over the ensemble of backgrounds.
There are actually two terms of this form, one coming from the charged
Higgs field background and the other from charged W bosons, but they
have the same general nature, so we lump them together.  
We will take the individual matrix elements $W_{IJ}$ to be of order one.
To understand the order of magnitude of the harmonic potentials , begin
with the special configuration where the matrix elements are equal to
one only between nearest neighbor points on the toroidal lattice of zero
branes and zero elsewhere.  Then the
frequency squared matrix for the set of coupled oscillators looks like
the Laplacian on a three dimensional toroidal lattice of side
$N^{(1/3)}$ .   The nonzero eigenvalues of this matrix are thus of order
$N^{-(2/3)}$.  The eigenvectors represent displacements of the zero
branes away from the center of mass of the system.  When these
displacements are of order the Schwarzchild radius , that is
$o(N^{1/6})$, these harmonic energies are of the same order as the
single particle energies of the zero branes computed in the mean field
approximation with their velocity dependent two body interactions.  
Thus, a consistent picture emerges, in which both harmonic attraction to
the background and interparticle forces bind the zero branes at a
distance of order the Schwarzchild radius from the center of mass.   

Small changes of individual $W_{IJ}$ matrix elements are equivalent to
changing the Laplacian on a flat toroidal lattice to that on a curved
background .  The inclusion of nonzero matrix elements between
nonnearest neighbors (but always a distance of order one away from each
other) adds higher derivative terms to the Laplacian and does not change
the qualitative scaling behavior for large $N$.  

Finally, we must check the terms in the energy coming from the square of
the commutator of charged fields.  Consider for example matrices whose
only nonzero matrix elements are between nearest neighbors on the
toroidal lattice.  If all of the nearest neighbor matrix elements were
equal, the matrix would be a sum of commuting shift operators on the
torus.  Thus, the commutator of two such
local matrices is
proportional to differences of nearest neighbor matrix elements.  The
Hamiltonian is proportional to the trace of the 
square of the commutator.  If
we take nearest neighbor matrix elements of order one, with differences
between them of order $N^{-(1/6)}$, then the energy will be a sum of $N$
terms, each of order $N^{-(1/3)}$.    Clearly, these estimates
generalize to matrices with nonzero matrix elements for $o(1)$ next to
nearest neighbors.
Thus, it is possible to find
classical configurations of charged fields whose energy and entropy are
of the same order as those of the zero branes.

It is interesting that the background entropy is so low in eleven
dimensions but appears to be a finite fraction of the total black hole
entropy in eight.  We do not have a clear picture of the significance of
this result, and we do not know how it generalizes to other dimensions.
In seven dimensions, although the compactified theory is poorly
understood, we believe that we have identified background configurations
involving tensionless strings, which have energy per zero brane of order
$N^{-(2/5)}$.  This suggests that high entropy backgrounds will be found
in all dimensions below nine.  However, we do not know how to compute
the harmonic restoring force in this case and the analysis is too
preliminary to be presented here.

If the background contributes a
finite fraction of both the energy and entropy of the black hole, it is
somewhat artificial to separate the system into zero branes plus
background. We suspect that this may be the general case when the
noncompact dimension is less than nine.  
Nonetheless, we have chosen to present the 
statistical mechanics of matrix black holes in terms of the zero brane
gas, because this makes it clear that there will be configurations
satisfying the black hole mass--entropy--radius relation for Matrix
Theory compactifications to any dimension greater than five.  Even in
eight dimensions, we can find backgrounds with harmonic potentials of
order $1/N$ and build black holes from them.  They are simply much less
numerous than the configurations studied in \bfks\ .  

It is in four and five uncompactified dimensions that our approach
really appears to run into trouble.  We have so far been unable to find
configurations which produce a harmonic potential with coefficient
smaller than $1/N$.  In $D$ dimensions,at the Schwarzchild radius,
 this gives an energy of order
$N^{{2\over {D-2}} - 1}$, which should be compared to the energy per
particle of the D0 brane gas $N^{- {2 \over {D-2}}}$.  The ratio of
energies is $N^{{6-D}\over {D-2}}$.
Thus, in four and five dimensions the harmonic potential dominates and
our picture breaks down.

Actually, in four dimensions, this discussion is altogether too naive.
Indeed it is completely unclear whether the entire matrix model
formalism makes sense in four dimensions.  In the matrix model we are
supposed to treat four dimensions by first compactifying to three and
then decompactifying the longitudinal direction by taking $N$ to
infinity.   But in three dimensions, Kaluza Klein modes have long range
dilaton fields which render their BPS charges undefined.  The energy of
a generic collection of BPS Kaluza-Klein modes appears to be infinite.  
We suspect that this is a deep property of M theory which implies that
compactifications below four dimensions only make sense in a
cosmological context.  A discrete light cone approach could only make
sense above four dimensions.
This is clearly not the time or place to discuss
such issues.  

This leaves us with five dimensional compactifications of the matrix
model. We are unsure whether the difficulty which our model encounters
indicates the necessity of finding special background configurations
with small harmonic potentials for zero branes, or a deeper sickness in
the five dimensional theory.  In the Appendix we suggest that
the sickness may be avoided in the regime $N\gg S$.

\newsec{\bf Conclusions}

We have been led to a remarkably simple picture of Schwarzchild 
black holes in matrix
theory, which appears to be valid when there are more than five
noncompact dimensions.  They are collections of zero branes, interacting
with themselves and with a classical background.  In eleven dimensions
the background appears to be necessary only to \lq\lq Boltzmannize\rq\rq\ 
the zero branes, that is to break their statistical gauge
symmetry. It does not contribute appreciably to the energy or entropy of
the black hole.  In eight dimensions, the background does make
substantial contributions to the entropy, and we expect that this may be
the general case with eight or fewer compact dimensions.

One obvious calculation which we have not done is the Hawking evaporation
rate.  Given the parallel between the kinematics of Hawking particles
and our D0 branes, a natural guess for the Hawking evaporation mechanism
is a quantum fluctuation which erases the part of the classical
background which interacts with a given D0 brane.  In our $8$
dimensional model for example, this consisted of $o(1)$ strings.  We
have not yet been able to estimate the probability for these strings to
disappear and we are not able to understand why it should decrease like a power
of $N$ (let alone compute the power).   If indeed this probability falls
like an appropriate power of $N$ then we could understand the Hawking
evaporation formula from the microscopic mechanics of our model.  This
is an interesting topic for further study.

Apart from the obvious call to solve the mysteries of low dimensional
compactification, our approach should be extended in various directions.
One should obtain a clear picture of the relevant background
configurations for all $D>5$.  One should also
extend our considerations to
charged and rotating black holes.   Perhaps most interesting of all
would be to model the experience of an \lq\lq observer\rq\rq\ falling
into our matrix black hole, and to extract the spacetime metric which he
feels.

Finally, we remark that 
recent work on scattering in Matrix Theory\dr\ \ref\dosdo{M.R.Douglas,
H.Ooguri, S.Shenker, Phys.Lett. B402 (1997) 36,hep-th/9702203; 
M.R.Douglas, H.Ooguri, hep-th/9710178. }
 has made it clear that the correct gravitational physics cannot be
 extracted from the matrix model at small values of $N$, unless it is
 protected by SUSY.  Although our
 analysis depended mostly on large $N$ scaling laws, we do not know the
 extent to which our considerations will be affected by these results.
 In particular, our estimate of the value of $N$ necessary to reproduce
 black hole physics, takes no account of the transverse separation of
 the black holes.  It may be that the correct physics can only be
 obtained in a regime in which the black hole wave functions overlap
 substantially.  In this case our derivation of the Newtonian
 interaction between black holes will appear accidental.   
We hope that this is not the case, since the Boltzmann gas model is
appealing in its simplicity.  The best way to attack these questions is
to see what other properties of black holes can be derived from the
model.

\centerline{\bf ACKNOWLEDGEMENTS}

T.B. would like to acknowledge conversations with D.Gross, J.Polchinski
and \break G.Horowitz, and the hospitality of ITP Santa Barbara, where those
conversations took place in the summer of 1997.  TB and WF would like to
thank the Physics Department of Stanford University for its hospitality
in July and August of 1997 when this work was begun.  W.Fischler would
like to thank the Physics Dept of Rutgers University for its hospitality
during the time that this work was completed.  All of the
authors would like to thank S.Shenker for collaborating on portions of
this work and for making his usual penetrating observations about all of
it. The work of TB was supported in part by the 
Department of Energy under Grant No. DE - FG02
- 96ER40959 and that  of W.F. was supported in part by the Robert A. Welch
Foundation and by NSF Grant PHY-9511632.
The work of I.R.K was supported in part by the DOE grant DE-FG02-91ER40671,
the NSF Presidential Young Investigator Award PHY-9157482, and the
James S.{} McDonnell Foundation grant No.{} 91-48.
L.S. acknowledges the support of the NSF under Grant No. PHY - 9219345.

\bigskip

\centerline{\bf Appendix: The Well Resolved Black Holes }

\bigskip

In section 5 we showed that our approach runs into
trouble for black holes in $D<6$.
Note that in another approach
\ref\ks{I.R.Klebanov, L.Susskind, hep-th/9709108. }
the search for black holes in matrix theory compactified to five
dimensions has led to the prediction
that the system has negative specific heat.  
This is not unexpected because $D=5$ is, of course, the
dimension where the effective theory of the matrix model (at least for
toroidal compactifications) seems to
contain gravitational degrees of freedom 
\ref\nati{N.Seiberg, hep-th/9710009.}.   
Let us note, however, that the pathologies with black holes have appeared in
the regime $N\sim S$ where there are barely enough degrees of freedom
to describe them qualitatively.
In this section we suggest that they may be avoided for $N\gg S$
where the black holes are well resolved and we may hope for a
quantitative description.

Let us see what form of the light-cone equation of state
 is necessary to reproduce the mass-entropy relation for Schwarzschild
black holes. Using $E= M^2 R/N$, we arrive at
\eqn\eos{ S \sim G_N^{1\over D-3} 
\left ( {NE\over R}\right )^{D-2\over 2(D-3)}
\ .}
The specific heat implied by this is positive for $D>4$.

$D=4$ is a special case where the specific heat is infinite,
$$ S \sim {N G_N\over R} E
\ .
$$
This behavior is characteristic of a gas of strings whose tension
scales as $1/N^2$. Strings of this kind have been conjectured to exist
in M-theory compactified on $T^7$ \ref\HK{A. Hanany and I.R. Klebanov,
Nucl. Phys. {\bf B482 } (1996) 105.}, but their relevance to this
equation of state is unclear.

For $D=5$ \eos\ gives $S\sim E^{3/4}$, which is characteristic of a
$3+1$ dimensional massless field theory.
In Matrix theory compactified on $T^6$ there are 3 transverse directions.
In \ref\sen{A. Sen, hep-th/9709220.} it was suggested that the 0-branes
become smeared in these directions, so that an appropriate description
is $3+1$ dimensional. Our comparison with Schwarzschild black holes
seems to
give an independent reason to believe that Matrix theory compactified on 
$T^6$ is described at very low energies by such a field theory.

For $D=6$, \eos\ gives $S\sim E^{2/3}$, which is characteristic of a
$2+1$ dimensional massless field theory.
For $D>6$ there are no cases where we find a field theoretic scaling,
$S\sim E^{p\over p+1}$. 

It is interesting that, only in cases where 
the specific heat is infinite or negative for $N\sim S$, do we find
recognizable scalings for $N\gg S$. Based on this we speculate that
the appropriate setting for black holes in $D=4$ and $5$ (and perhaps
even in $D=6$) is to work with $N\gg S$.

\listrefs
\end